\documentclass{iau}
\usepackage{graphicx}
\usepackage{natbib}

\title[Pulsar glitches and their impact on neutron-star astrophysics] 
{Pulsar Glitches}

\author[R. N. Manchester]   
{R. N. Manchester}

\affiliation{CSIRO Astronomy and Space Science, Sydney, Australia \\ email: {\tt dick.manchester@csiro.au}}

\pubyear{2017}
\volume{337}  
\jname{Pulsar Astrophysics -­ The Next 50 Years}
\editors{P. Weltevrede, B.B.P. Perera, L. Levin Preston \& S. Sanidas, eds.}

\begin{document}

\maketitle

\begin{abstract}
The first known pulsar glitch was discovered in the Vela pulsar at
both Parkes and Goldstone in March 1969. Since then the
number of known glitches has grown enormously, with more than 520
glitches now known in more than 180 pulsars. Details of glitch
parameters and post-glitch recoveries are described and some
implications for the physics of neutron stars are discussed.

\end{abstract}

\firstsection 
\section{The discovery of glitches and the starquake model}
\begin{figure}[b]
  \begin{center}
    \begin{minipage}{138mm}
      \begin{tabular}{cc}
        \mbox{\includegraphics[height=60mm]{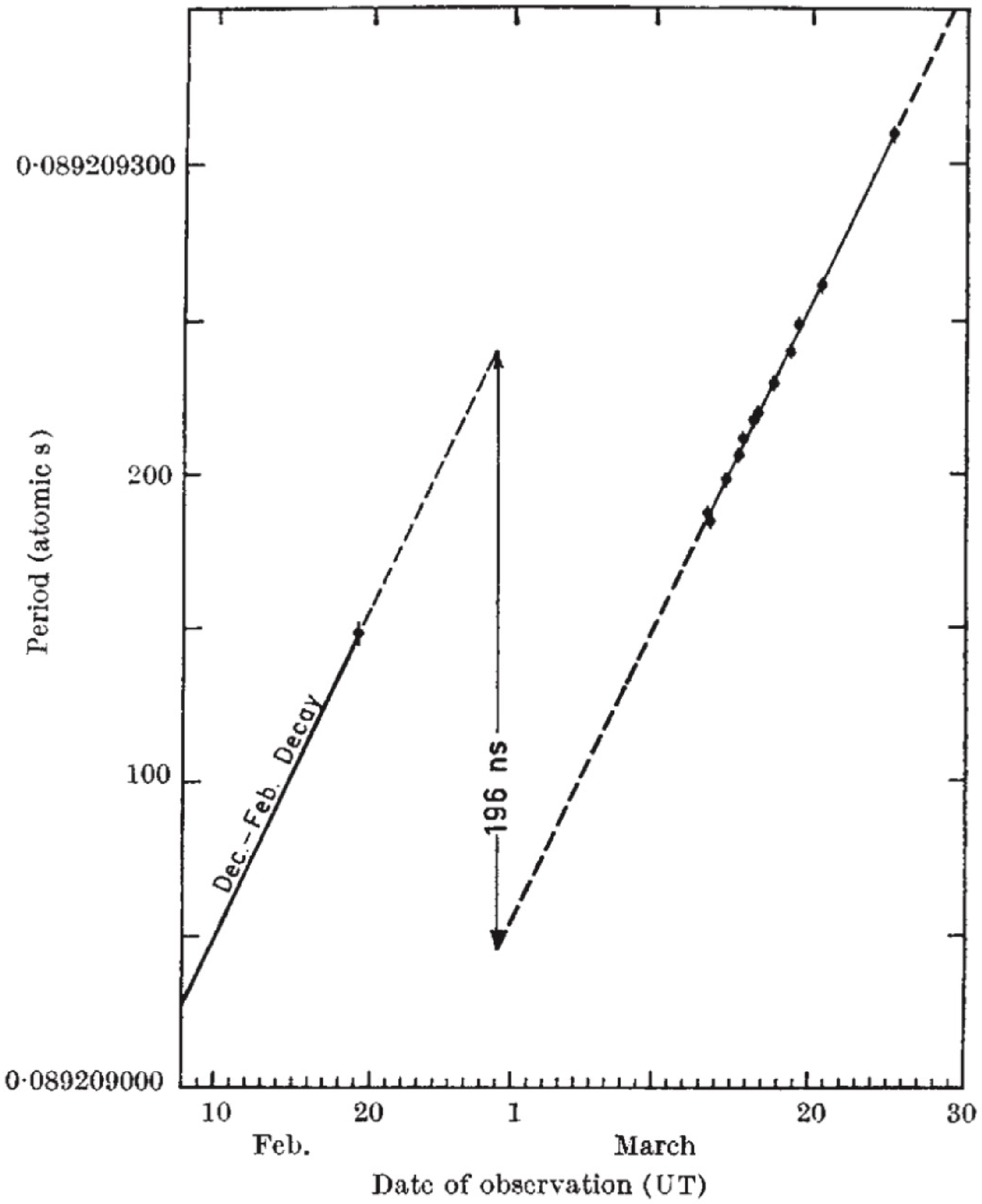}} &
        \mbox{\includegraphics[height=60mm]{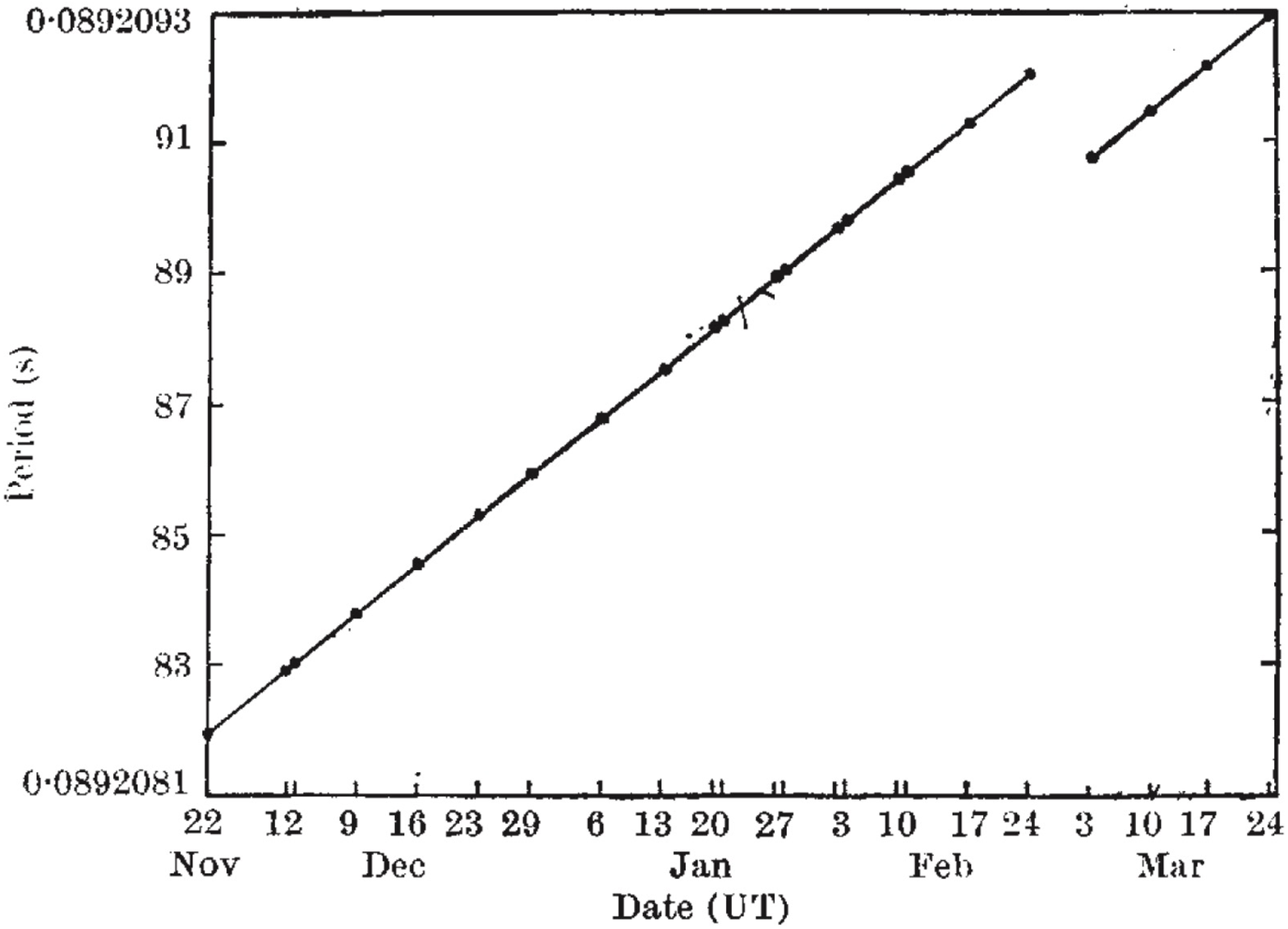}}
      \end{tabular}
      \caption{Changes in the Vela pulsar period in late 1968
        and early 1969 showing the first detections of a pulsar
        glitch. The left panel shows the Parkes observations
        \citep{rm69} and the right panel shows the JPL Goldstone
        observations \citep{rd69}.}\label{fg:vela-gl}
    \end{minipage}
  \end{center}
\end{figure}

{\let\thefootnote\relax\footnote{{$^{\dagger}$The term ``glitch'' was not initally used to describe
  these events. The first published use of the term appears to be by
  \citet{rtc71}.}}}
In late 1968 and early 1969, I was at Parkes and, among other things,
helping Radhakrishnan with the observations of the Vela pulsar that
ultimately led to the ``rotating vector model'' for pulsar
polarisation \citep{rc69a}. In mid-March, 1969, we set up the
signal-averager to fold the Vela pulsar data at the predicted
topocentric period and noticed that the pulse was slowly drifting
backwards on the screen, indicating that the folding period was not
quite correct. After exhaustively checking the equipment and observing
other pulsars, we concluded that everything was working correctly and,
consequently, that the pulsar period $P$ was not as predicted. The
implied period decrease $\Delta P$ was 196~ns, corresponding to a
relative change $\Delta\nu/\nu = -\Delta P/P \sim 2.2\times 10^{-6}$,
where the pulsar rotation frequency $\nu = 1/P$. We contacted Paul
Reichley and George Downs, whom we knew were timing the Vela pulsar
using the Goldstone antenna of the Jet Propulsion Laboratory (JPL) in
California, and confirmed that they also had seen the
glitch (Figure~\ref{fg:vela-gl}).$^\dagger$ Back-to-back papers reporting the discovery were
published in the April 19 Nature \citep{rm69,rd69}. The JPL
observations limited the glitch epoch to between February 24 and March
3, and also revealed a change in the slow-down rate
$\Delta\dot\nu/\dot\nu = \Delta\dot P/\dot P \sim 10^{-2}$. Both
groups suggested a sudden decrease in the neutron-star moment of
inertia, which could account for the changes in both $\nu$ and
$\dot\nu$. The required effective change in the radius of the neutron
star was about 1~cm.

Within a few months, \citet{bppr69} had refined this ``starquake''
model, suggesting that the change in moment of inertia was due to
relaxation of the neutron-star crust to the current equilibrium
oblateness, which of course changes because of the gradual slow-down
in rotation. They also predicted that the observed increase in
$|\dot\nu|$ would decay on a timescale of years because of the weak
frictional coupling of a more rapidly rotating superfluid interior,
earlier predicted to exist by Russian theorists \citep{mig60,gk65},
and the neutron-star crust to which the emission beams are locked.

A prediction of this model was that it would be at least 300 years
before the stresses due to the oblateness differential built up sufficiently
to cause another starquake. The JPL group continued their regular
monitoring and, just 2.5 years after the first glitch, announced the
detection of a second large glitch with $\Delta\nu/\nu \sim 2.0\times
10^{-6}$ \citep{rd71}. This of course ruled out relaxation of crustal
oblateness as a mechanism for the glitch trigger. Alternative models
quickly followed, with ``corequakes'' suggested by \citet{psr72} and
the sudden unpinning of interior superfluid vortices, with a
consequent transfer of angular momentum to the crust, first suggested by
\citet{ai75}. As will be discussed in Section~\ref{sec:interp} below, the
latter idea forms the basis for most subsequent interpretations of the
glitch phenomenon.

Both \citet{rm69} and \citet{rd69} pointed out in their concluding
remarks that glitches could be expected in the Crab pulsar
period. Sure enough, about six months later, \citet{bgpw69} and
\citet{rpr+69} announced the discovery of a glitch in the Crab pulsar
period. The relative glitch size, $\Delta\nu/\nu \sim 7\times
10^{-9}$, was about 300 times smaller than for the Vela glitches
suggesting a different mechanism. Observations over the next
few years at Jodrell Bank and optical observatories
\citep[e.g.,][]{loh81,lps93} revealed several glitches, including
larger ones in 1975 and 1989. These observations also showed
that the post-glitch behaviour in the Crab pulsar was quite different
to that for Vela, being dominated by a persistent increase
in the slow-down rate $|\dot\nu|$.

\section{The glitch population}\label{sec:pop}
\begin{figure}[b]
  \begin{center}
    \includegraphics[height=95mm]{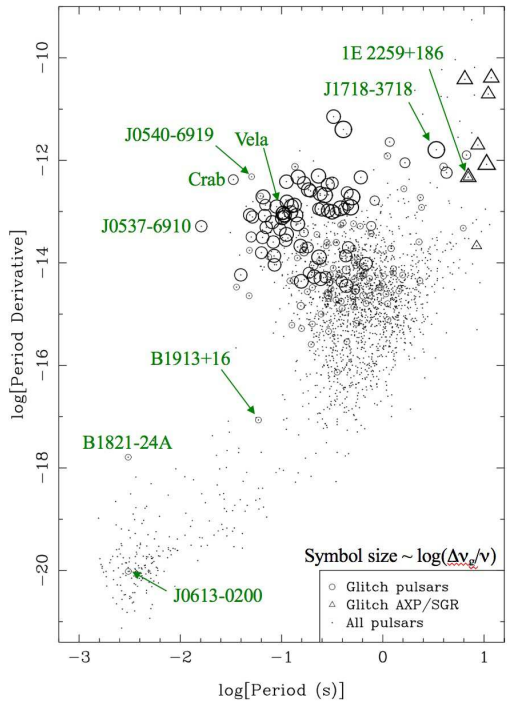}
    \caption{Distribution of the 180 glitching pulsars on the $P$ --
      $\dot P$ diagram. The symbol size is proportional to the
      logarithm of the largest fractional glitch size $\Delta\nu/\nu$
      observed for each pulsar. Glitching pulsars and AXPs/SGRs
      (magnetars) and non-glitching pulsars are marked with different
      symbols as labelled. }\label{fg:gl_ppdot}
  \end{center}
\end{figure}

{\let\thefootnote\relax\footnote{{$^{\ast}$http://www.jb.man.ac.uk/pulsar/glitches/gTable.html}}}
{\let\thefootnote\relax\footnote{{$^{\ddagger}$http://www.atnf.csiro.au/research/pulsar/psrcat/glitchTbl.html}}}
Tables of observed glitches are maintained by Jodrell Bank Observatory
(JBO)$^\ast$ and as part of the ATNF Pulsar Catalogue$^\ddag$. While
these two tables broadly contain the same information, there is some
information in one, but not the other. In particular, the JBO table
contains details of about 80 otherwise unpublished glitches. Collating
the data from the two tables gives a total of 520 known glitches in
180 different pulsars. Figure~\ref{fg:gl_ppdot} shows the distribution
of glitching pulsars on the $P$ -- $\dot P$ diagram. This figure shows
that glitches in young pulsars, including magnetars, are generally
large with $\Delta\nu/\nu \sim 10^{-6}$. However, the youngest
pulsars, e.g., the Crab pulsar, PSR J0537$-$6910 and PSR J0540$-$6919,
tend to have more frequent and smaller glitches with $\Delta\nu/\nu
\sim 10^{-7}$ or $10^{-8}$. The most frequently glitching pulsar is
PSR J0537$-$6910, located in the Large Magellanic Cloud which, on
average, glitches about three times per year \citep{mgm+04}.  Only two
millisecond pulsars, PSRs B1821$-$24A and J0613$-$0200, have been
observed to glitch and each of these have had just one very small
glitch with $\Delta\nu/\nu \sim 10^{-11}$ \citep{cb04,mjs+16}. It is
interesting that the Hulse -- Taylor binary pulsar, PSR B1913+16, also
had a small glitch in 2003 of about the same magnitude \citep{wnt10}.

Large glitches are relatively common in magnetars. While their
magnitudes are similar to those in other young pulsars, they have a
number of distinguishing features. For example, they are sometimes
accompanied by radiative changes, either X-ray bursts or associated
changes in the pulse profile \citep[see, e.g.,][]{dk14}. Such
X-ray bursts and profile changes are very common in magnetars and
are generally accompanied by timing irregularities, but
only a small proportion of them are associated with glitches. Since
magnetar glitch properties are broadly similar to those in other young
pulsars, it seems plausible that the glitch mechanism is
the same or similar, i.e., related to changes in the superfluid
interior of the star. In this case, the radiative associations suggest
some connection between the stellar interior and the magnetosphere of
the star.

However, \citet{akn+13} reported the detection of an ``anti-glitch'', i.e., an
abrupt spin-down, in the period of the magnetar 1E~2259+586 (PSR
J2301+5852) of relative magnitude about $3.1\times 10^{-7}$. Several
large glitches, one with $\Delta\nu/\nu \sim 1.6\times 10^{-5}$, have
also been seen in this pulsar \citep{dkg09}. The anti-glitch was
associated with a short-duration hard X-ray burst and an increase in
the soft X-ray flux which decayed over 100 days or so. These
properties suggest that the anti-glitch was magnetospheric in origin.

Although glitches in normal young pulsars are generally not associated
with radiative changes, glitches observed in the very young pulsar PSR
J1119$-$6127 are exceptions. \citet{wje11} observed that, for about
three months after a large glitch ($\Delta\nu/\nu \sim 1.6\times
10^{-6}$), intermittent strong pulses and a second profile component
were observed. An even larger subsequent glitch ($\Delta\nu/\nu \sim
5.7\times 10^{-5}$) was observed by \citet{akts16} to be accompanied
by X-ray bursts and X-ray pulsations. It seems as though this pulsar
is half-way to being a magnetar.

\section{Glitch properties and their interpretation}\label{sec:interp}
Glitch activity in pulsars can be quantified by the relation
\begin{equation}
  A_g = \frac{1}{T}\frac{\Sigma(\Delta\nu_g)}{\nu}
\end{equation}
where $T$ is the total data span for glitch monitoring and
$\Sigma(\Delta\nu_g)$ is the sum of all observed glitch frequency
jumps. These jumps reverse some fraction of the regular slow-down due
to electromagnetic and wind torques. Many studies \citep[see,
  e.g.,][]{elsk11,ymh+13} have shown that for pulsars with
characteristic ages $\tau_c = P/(2\dot P)$ between about $10^4$ and
$10^5$ years, about 1\% of the slowdown is reversed by glitches. For
the Crab and other very young pulsars, glitches have about two orders
of magnitude less effect on the slow-down. In the two-component
superfluid models \citep[e.g.,][]{aaps81}, the glitch results from the
sudden unpinning and then repinning of vortex lines transferring
angular momentum from the more rapidly spinning interior superfluid to
the crust. The moment of intertia of the pinning/unpinning superfluid
$I_s$ is related to the glitch activity by the ``coupling parameter''
\begin{equation}
  G = 2\tau_c A_g = \frac{\dot\nu_g}{|\dot\nu|} \sim \frac{I_s}{I}
\end{equation}
where I is the total moment of inertia of the neutron star.

\begin{figure}[b]
  \begin{center}
    \begin{minipage}{138mm}
      \begin{tabular}{cc}
        \mbox{\includegraphics[height=80mm]{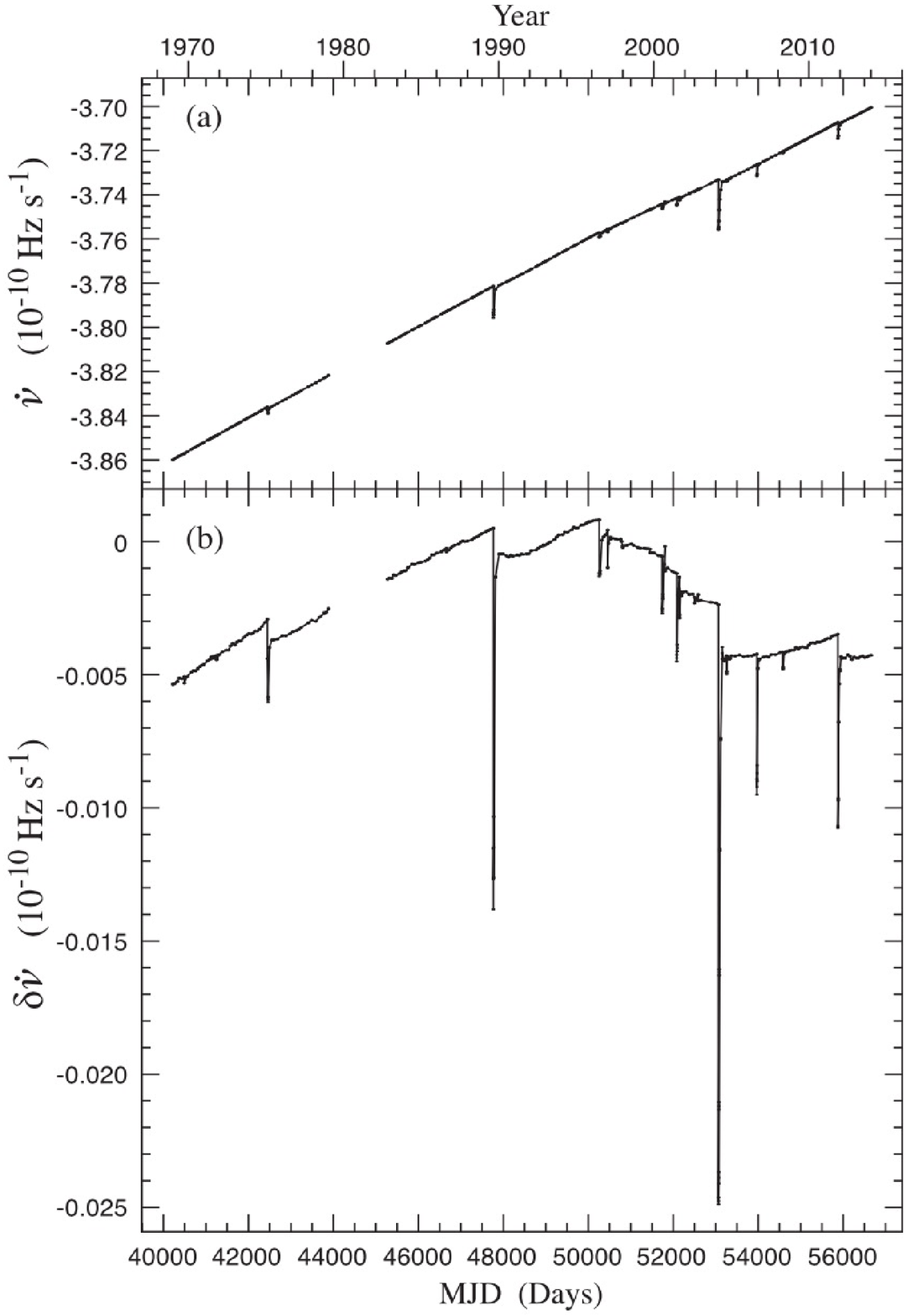}} &
        \mbox{\includegraphics[height=80mm]{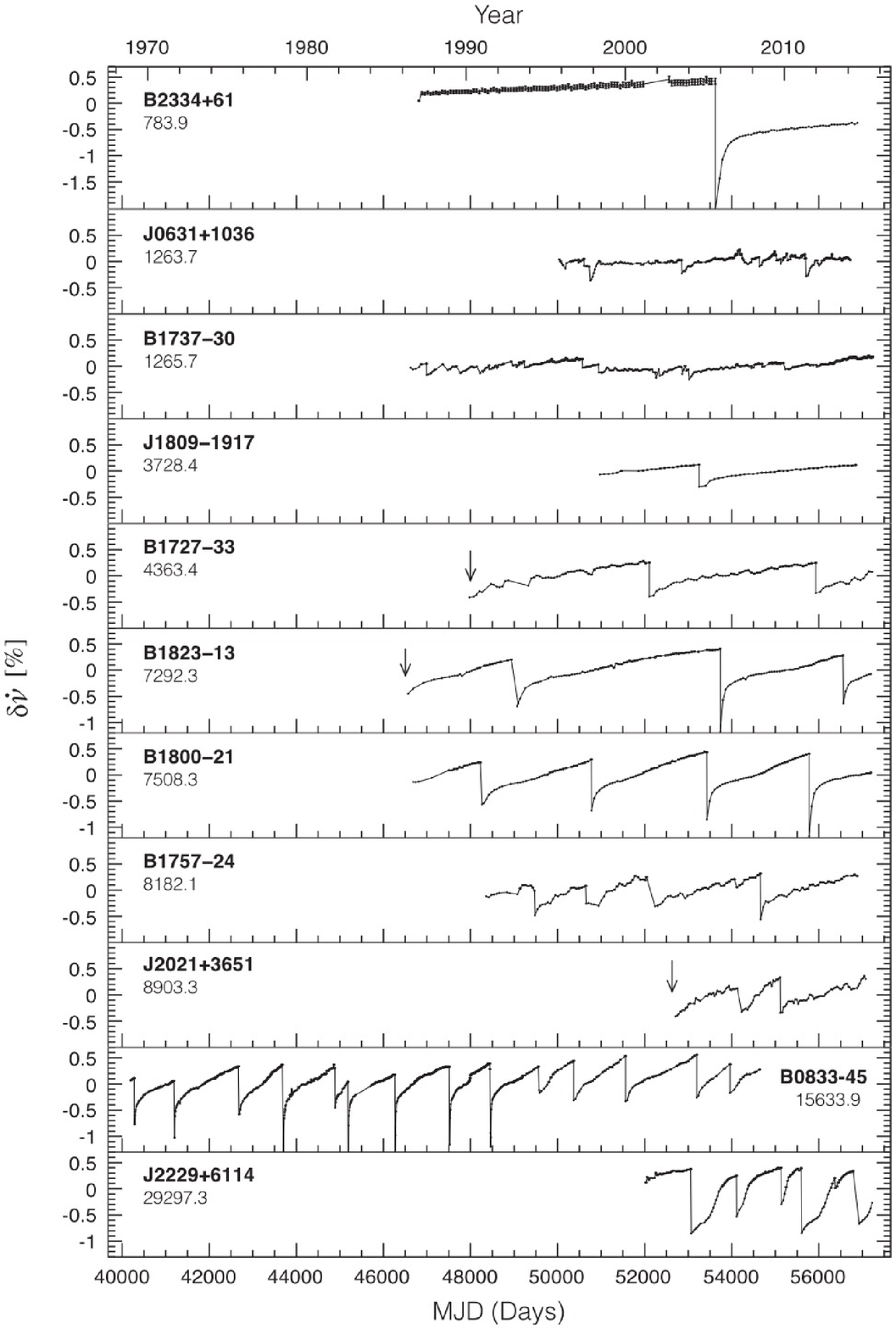}}
      \end{tabular}
      \caption{Variations of spin-down rate $\dot\nu$ for the Crab
        pulsar \citep[left,][]{ljg+15} and a sample of young pulsars
        \citep[right,][]{els17}). For the Crab pulsar, the lower
        subpanel shows the $\dot\nu$ variations after subtraction of
        the linear trend evident in the upper subpanel.}\label{fg:postgl}
    \end{minipage}
  \end{center}
\end{figure}

Many different types of post-glitch behaviour are observed. In some
pulsars, most or all of the initial frequency jump decays, whereas in
other cases the glitch is like a step jump in frequency with little or
no change in $\dot\nu$
\citep[e.g.,][]{elsk11,ymh+13}. Figure~\ref{fg:postgl} shows the
observed long-term variations in $\dot\nu$ for the Crab and a number
of other young pulsars. Glitches are marked by a sudden decrease in
$\dot\nu$. The fractional increase in slow-down rate $|\dot\nu|$ is typically
about 1\% although for some pulsars the increase is much smaller,
even not detectable. For most pulsars, much of this initial increase
decays exponentially on a timescale of 10 -- 100 days. For the Crab
pulsar, about 90\% of the increase quickly decays, but the remaining
10\% persists as a long-term increase in the slow-down rate. For large
glitches in other pulsars, typically about half of the initial
increase decays exponentially. Following that, there is a basically
linear increase in $\dot\nu$ until the next glitch.

The simple two-component model of \citet{bppr69} cannot account for
these different post-glitch behaviours. In a series of papers, Ali
Alpar and his colleagues have developed this model with different
regions within the neutron star having different properties to account
for the different post-glitch behaviours
\citep[e.g.,][]{aaps81,accp93,accp96}. For example, in regions
with weakly pinned vortices, vortex creep can occur, whereas in
strongly pinned regions, there is no creep following the repinning of
vortices after a glitch. Weakly pinned regions have a linear dynamical
response and can give the observed exponential recoveries in
$\dot\nu$. More strongly pinned regions have a non-linear dynamical
response that can result in a long-term linear increase in $\dot\nu$
as observed for Vela and other young pulsars. 

While the Alpar et al. models have been broadly successful in
accounting for the properties of pulsar glitches, they depend on
many assumptions about poorly known properties of neutron star
interiors. Various authors have proposed alternative views about some
of these assumptions. For example, \citet{cha12} has argued that
``entrainment'' of the neutron superfluid by the crystalline structure
of the crust greatly reduces its mobility. Consequently, unpinning
of the crustal superfluid is insufficient to account for large
glitches, and other mechanisms, e.g., unpinning of core superfluid
neutrons, are required. However, in a recent paper, \citet{wp17} argue
that entrainment is not a significant issue and there is no need to
invoke core superfluid. In another recent paper, \citet{lin14} has argued that the
``linear-response'' regime invoked by Alpar et al. is strongly
suppressed by a high vortex activation energy. If this is the case,
the interpretation of the exponential post-glitch decays invoked by,
e.g., \citet{accp93} would not be viable.

Various authors have used observed glitch properties as a probe to
investigate the mass of neutron stars. For example, \citet{pahs17}
used the largest observed glitch in a given pulsar to limit the
neutron-star mass on the assumption that there was complete unpinning
of vortices at a glitch, transferring the entire excess angular
momentum to the crust. They gave mass estimates for all pulsars in
which at least two large glitches had been observed. For Vela and PSR
J0537$-$6910 in particular, they obtained mass estimates of $1.35\pm
0.08$~M$_\odot$ and $1.25\pm 0.06$~M$_\odot$, respectively. In
contrast, \citet{heaa17} used detailed modelling of neutron-star
thermal evolution combined with nuclear equation-of-state and
superfluid models to interpret the size and rate of observed glitches
as a function of the neutron-star mass. For the same two pulsars,
\citet{heaa17} obtained mass estimates of $1.51\pm 0.04$~M$_\odot$ and
$1.83\pm 0.04$~M$_\odot$, respectively. The differences between these
derived masses clearly indicate that uncertainties in the physical
properties of neutron-star interiors lead to large systematic offsets
in estimated neutron-star masses. A positive aspect of this is that
pulsar glitches can provide useful input into the determination of
these properties.

\section*{Acknowlegements}
I thank the organisers of this Symposium for inviting me to review
this topic and my colleagues with whom I have had many fruitful
discussions and collaborations over the past five decades or so.


\begin{thebibliography}{37}
\expandafter\ifx\csname natexlab\endcsname\relax\def\natexlab#1{#1}\fi

\bibitem[{Alpar {et~al.}(1981)Alpar, Anderson, Pines, \& Shaham}]{aaps81}
Alpar, M.~A., Anderson, P.~W., Pines, D., \& Shaham, J. 1981, \textit{Astrophys. J.}, 249, L29

\bibitem[{{Alpar} \& {Baykal}(2006)}]{ab06}
{Alpar}, M.~A. \& {Baykal}, A. 2006, \textit{MNRAS}, 372, 489

\bibitem[{Alpar {et~al.}(1993)Alpar, Chau, Cheng, \& Pines}]{accp93}
Alpar, M.~A., Chau, H.~F., Cheng, K.~S., \& Pines, D. 1993, \textit{Astrophys. J.}, 409, 345

\bibitem[{Alpar {et~al.}(1996)Alpar, Chau, Cheng, \& Pines}]{accp96}
---. 1996, \textit{Astrophys. J.}, 459, 706

\bibitem[{Anderson \& Itoh(1975)}]{ai75}
Anderson, P.~W. \& Itoh, N. 1975, \textit{Nature}, 256, 25

\bibitem[{{Archibald} {et~al.}(2013){Archibald}, {Kaspi}, {Ng},
    {Gourgouliatos}, {Tsang}, {Scholz}, {Beardmore}, {Gehrels}, \&
    {Kennea}}]{akn+13}
  {Archibald}, R.~F., {Kaspi}, V.~M., {Ng},
  C.-Y., et al. 2013, \textit{Nature}, 497, 591

\bibitem[{{Archibald} {et~al.}(2016){Archibald}, {Kaspi}, {Tendulkar}, \&
  {Scholz}}]{akts16}
{Archibald}, R.~F., {Kaspi}, V.~M., {Tendulkar}, S.~P., \& {Scholz}, P. 2016,
  \textit{Astrophys. J.}, 829, L21

\bibitem[{Baym {et~al.}(1969)Baym, Pethick, Pines, \& Ruderman}]{bppr69}
Baym, G., Pethick, C., Pines, D., \& Ruderman, M. 1969, \textit{Nature}, 224, 872

\bibitem[{Boynton {et~al.}(1969)Boynton, Groth, Partridge, \&
  Wilkinson}]{bgpw69}
Boynton, P.~E., Groth, E.~J., Partridge, R.~B., \& Wilkinson, D.~T. 1969, \textit{IAU
  Circ.} 2179

\bibitem[{{Chamel}(2012)}]{cha12}
{Chamel}, N. 2012, \textit{Phys. Rev. C}, 85, 035801

\bibitem[{{Cognard} \& {Backer}(2004)}]{cb04}
{Cognard}, I. \& {Backer}, D.~C. 2004, \textit{Astrophys. J.}, 612, L125

\bibitem[{{Dib} \& {Kaspi}(2014)}]{dk14}
{Dib}, R. \& {Kaspi}, V.~M. 2014, \textit{Astrophys. J.}, 784, 37

\bibitem[{{Dib} {et~al.}(2009){Dib}, {Kaspi}, \& {Gavriil}}]{dkg09}
{Dib}, R., {Kaspi}, V.~M., \& {Gavriil}, F.~P. 2009, \textit{Astrophys. J.}, 702, 614

\bibitem[{{Espinoza} {et~al.}(2017){Espinoza}, {Lyne}, \& {Stappers}}]{els17}
{Espinoza}, C.~M., {Lyne}, A.~G., \& {Stappers}, B.~W. 2017, \textit{MNRAS}, 466, 147

\bibitem[{{Espinoza} {et~al.}(2011){Espinoza}, {Lyne}, {Stappers}, \&
  {Kramer}}]{elsk11}
{Espinoza}, C.~M., {Lyne}, A.~G., {Stappers}, B.~W., \& {Kramer}, M. 2011,
  \textit{MNRAS}, 414, 1679

\bibitem[{{Ginzburg} \& {Kirzhnits}(1965)}]{gk65}
{Ginzburg}, V.~L. \& {Kirzhnits}, D.~A. 1965, \textit{Soviet Physics JEPT}, 20, 1346

\bibitem[{{Ho} {et~al.}(2017){Ho}, {Espinoza}, {Antonopoulou}, \&
  {Andersson}}]{heaa17}
{Ho}, W.~C.~G., {Espinoza}, C.~M., {Antonopoulou}, D., \& {Andersson}, N. 2017,
  \textit{14th Int. Symp. Nuclei in the Cosmos}, 010805

\bibitem[{{Link}(2014)}]{lin14}
{Link}, B. 2014, \textit{Astrophys. J.}, 789, 141

\bibitem[{{Link} {et~al.}(1998){Link}, {Franco}, \& {Epstein}}]{lfe98}
{Link}, B., {Franco}, L.~M., \& {Epstein}, R.~I. 1998, \textit{Astrophys. J.}, 508, 838

\bibitem[{Lohsen(1981)}]{loh81}
Lohsen, E. H.~G. 1981, \textit{Astron. Astrophys. Suppl.}, 44, 1

\bibitem[{{Lyne} {et~al.}(2015){Lyne}, {Jordan}, {Graham-Smith}, {Espinoza},
  {Stappers}, \& {Weltevrede}}]{ljg+15}
{Lyne}, A.~G., {Jordan}, C.~A., {Graham-Smith}, F., {Espinoza}, C.~M.,
  et al. 2015, \textit{MNRAS}, 446, 857

\bibitem[{Lyne {et~al.}(1993)Lyne, Pritchard, \& Smith}]{lps93}
Lyne, A.~G., Pritchard, R.~S., \& Smith, F.~G. 1993, \textit{MNRAS}, 265, 1003

\bibitem[{{Marshall} {et~al.}(2004){Marshall}, {Gotthelf}, {Middleditch},
  {Wang}, \& {Zhang}}]{mgm+04}
{Marshall}, F.~E., {Gotthelf}, E.~V., {Middleditch}, J., et al. 2004, \textit{Astrophys. J.}, 603, 682

\bibitem[{{McKee} {et~al.}(2016){McKee}, {Janssen}, {Stappers}, {Lyne},
  {Caballero}, {Lentati}, {Desvignes}, {Jessner}, {Jordan}, {Karuppusamy},
  {Kramer}, {Cognard}, {Champion}, {Graikou}, {Lazarus}, {Os{\l}owski},
  {Perrodin}, {Shaifullah}, {Tiburzi}, \& {Verbiest}}]{mjs+16}
{McKee}, J.~W., {Janssen}, G.~H., {Stappers}, B.~W., et al. 2016, \textit{MNRAS}, 461,
  2809

\bibitem[{{Migdal}(1960)}]{mig60}
{Migdal}, A.~B. 1960, \textit{Soviet Physics JEPT}, 10, 176

\bibitem[{Pines {et~al.}(1972)Pines, Shaham, \& Ruderman}]{psr72}
Pines, D., Shaham, J., \& Ruderman, M.~A. 1972, \textit{Nature Phys. Sci.}, 237, 83

\bibitem[{{Pizzochero} {et~al.}(2017){Pizzochero}, {Antonelli}, {Haskell}, \&
  {Seveso}}]{pahs17}
{Pizzochero}, P.~M., {Antonelli}, M., {Haskell}, B., \& {Seveso}, S. 2017,
  \textit{Nature Astronomy}, 1, 0134

\bibitem[{Radhakrishnan \& Cooke(1969)}]{rc69a}
Radhakrishnan, V. \& Cooke, D.~J. 1969, \textit{Astrophys. Lett.}, 3, 225

\bibitem[{Radhakrishnan \& Manchester(1969)}]{rm69}
Radhakrishnan, V. \& Manchester, R.~N. 1969, \textit{Nature}, 222, 228

\bibitem[{Reichley \& Downs(1969)}]{rd69}
Reichley, P.~E. \& Downs, G.~S. 1969, \textit{Nature}, 222, 229

\bibitem[{Reichley \& Downs(1971)}]{rd71}
---. 1971, \textit{Nature Phys. Sci.}, 234, 48

\bibitem[{{Richards} {et~al.}(1969){Richards}, {Pettengill}, {Roberts},
  {Counselman}, \& {Rankin}}]{rpr+69}
{Richards}, D.~W., {Pettengill}, G.~H., {Roberts}, J.~A., et al. 1969, \textit{IAU Circ.} 2181

\bibitem[{{Rees} {et~al.} (1971)}]{rtc71}
Rees, M.~J, Trimble, V.~L. \& Cohen, J.~M. 1971, \textit{Nature}, 229, 395

\bibitem[{{Watanabe} \& {Pethick}(2017)}]{wp17}
{Watanabe}, G. \& {Pethick}, C.~J. 2017, \textit{Phys. Rev. Lett.}, 119, 062701

\bibitem[{{Weisberg} {et~al.}(2010){Weisberg}, {Nice}, \& {Taylor}}]{wnt10}
{Weisberg}, J.~M., {Nice}, D.~J., \& {Taylor}, J.~H. 2010, \textit{Astrophys. J.}, 722, 1030

\bibitem[{{Weltevrede} {et~al.}(2011){Weltevrede}, {Johnston}, \&
  {Espinoza}}]{wje11}
{Weltevrede}, P., {Johnston}, S., \& {Espinoza}, C.~M. 2011, \textit{MNRAS}, 411, 1917

\bibitem[{{Yu} {et~al.}(2013){Yu}, {Manchester}, {Hobbs}, {Johnston}, {Kaspi},
  {Keith}, {Lyne}, {Qiao}, {Ravi}, {Sarkissian}, {Shannon}, \& {Xu}}]{ymh+13}
{Yu}, M., {Manchester}, R.~N., {Hobbs}, G., et al. 2013, \textit{MNRAS}, 429, 688

\end{thebibliography}




\end{document}